\documentclass[prl,twocolumn,superscriptaddress,showpacs,preprintnumbers]{revtex4}
\usepackage{graphicx}% Include figure files
\usepackage{dcolumn}% Align table columns on decimal point
\usepackage{bm}% bold math
\usepackage{multirow,amssymb,amsbsy,amsmath}

\newcommand{\I}{\textup{i}}

\newcommand{\ddim}{\udelta\kern0.1em}

\newcommand{\beikonst}[2]{\left( #1 \right)_{\kern-0.2em #2}}

\newcommand*{\ket}[1]{\mathopen{|}#1\mathclose{\rangle}}

\newcommand{\C}[1]{$^{#1}$C}
\begin{document}
\preprint{APS/123-QED}

% -----------------------------------------
%
% Titel
\title{Coherence of single spins coupled to a nuclear spin bath of varying density}

\author{N. Mizuochi$^{1}$, P. Neumann$^2$, F. Rempp$^2$, J. Beck$^2$, V. Jacques$^2$, P. Siyushev$^2$, K. Nakamura$^3$, 
D. Twitchen$^4$, H. Watanabe$^5$, S. Yamasaki$^6$, F. Jelezko$^2$, J. Wrachtrup$^2$}
\affiliation{
\mbox{$^1$Graduate School of Lib., Inf.\ \& Media Studies, University of Tsukuba, 305-8550 and PRESTO, JST, Japan}\\
\mbox{$^2$3.\ Physikalisches Institut, Universit\"at Stuttgart, Pfaffenwaldring 57, D-70550 Stuttgart, Germany}\\
\mbox{$^3$Tokyo Gas Co., Ltd., 3-13-1 Minamisenju, Tokyo, 116-0003, Japan}\\
\mbox{$^4$Element Six Ltd, King's Ride Park, Ascot, Berkshire SL5 8BP, UK}\\
\mbox{$^5$Diamond Research Center, AIST, Tsukuba Central 2, Tsukuba, 305-8568, Japan}\\
\mbox{$^6$Nanotechnology Research Institute AIST, Tsukuba Central 2, Tsukuba, 305-8568, Japan}%
}%
\received{\today}%

\begin{abstract}
The dynamics of single electron and nuclear spins in a diamond lattice with different \C{13} nuclear spin 
concentration is investigated. 
It is shown that coherent control of up to three individual nuclei in a dense nuclear spin cluster is feasible. 
The free induction decays of nuclear spin Bell states and single nuclear coherences among \C{13} nuclear spins 
are compared and analyzed. 
Reduction of a free induction decay time $T_2^*$ and a coherence time $T_2$ upon increase of nuclear spin 
concentration has been found. 
For diamond material with depleted concentration of nuclear spin, $T_2^*$ as long as 30 $\mu$s and $T_2$ of up to 
1.8 ms for the electron spin has been observed. 
The \C{13} concentration dependence of $T_2^*$ is explained by Fermi contact and dipolar interactions with nuclei 
in the lattice. 
It has been found that $T_2$ decreases approximately as 1/$n$, where $n$ is \C{13} concentration, as expected 
for an electron spin interacting with a nuclear spin bath.
\end{abstract}

%\pacs{03.65.-w, 03.67.-a, 03.67.Lx, 05.30.-d}% PACS, the Physics and Astronomy
                             % Classification Scheme.
\maketitle

% ---------------------------------------------------------------------------
%
% Main text
%
% ---------------------------------------------------------------------------
Defect centers in diamond have attracted considerable interest recently owing to their application for quantum 
information processing, communication and metrology.\ \cite{Fed,Dut,Neu,Jia,Maz,Gop,Ma2}
Especially the nitrogen-vacancy (NV) center, with its strong and spin dependant optical transitions allows for 
single spin readout and exquisite coherent control which is crucial for quantum information applications. \cite{Fed,Dut,Neu,Jia,Maz}
Owing to the high Debye temperature of diamond and weak coupling to acoustic phonons NV electron spins show 
long coherence time.
It was e.g. proposed to build small quantum registers by exploiting the interaction between the electron spin and 
a small number of nuclear spins in the immediate vicinity.
Five-qubit would be sufficient to perform all functions necessary for a node in a defect center based quantum 
repeater node.\ \cite{Jia,Maz} 
Up to now coherent control, swapping of coherence and even entanglement between up to two nuclei and the 
electron spin was demonstrated.\ \cite{Neu} 
To increase the size of the quantum register, more nuclei need to be coupled to the electron spin. 
The approach taken here is to increase the concentration of paramagnetic \C{13} nuclei in the lattice. 
We systematically demonstrate coherent control of up to three nuclear spins being coupled to an NV center electron spin in \C{13} 
isotopically enriched crystals, notwithstanding the fact that the electron decoherence time $T_2$ linearly scales 
with the \C{13} concentration. 
Furthermore, our experiments provide experimental insight into long studied problem of single central spin 
coupled to a paramagnetic environment.\ \cite{Chi,Tak,Han}
Owing to possibility to address individual electron spins in matrix with adjustable nuclear spin content we show 
the transition from diluted to dense spin bath (the situation relevant for spins in GaAs quantum dots).

\begin{figure}
\includegraphics{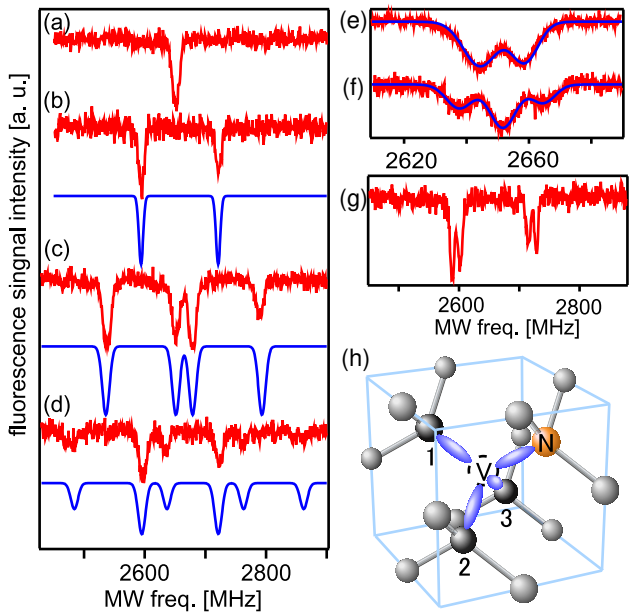}% Here is how to import EPS art[scale=0.48]
\caption{\label{ }(Color online) 
 ESR spectra of single NVs with (a) zero, (b) one, (c) two, (d) three \C{13} in the 1st shell. 
 Blue solid lines are simulation spectra. 
 ESR spectra with (e) one and (f) two \C{13} in the 3rd shell, and (g) one \C{13} in the 1st shell and one 
 \C{13} in the 3rd shell. Blue solid lines are fitting curves with Gaussian shape. 
 (h) Atomic structure of NV center. The numbers 1, 2, and 3 mark C in the 1st shell.}
\end{figure}

The quantum system used in the present work is the negatively charged NV center in diamond, 
which comprises a substitutional nitrogen atom with an adjacent vacancy. (Fig.\ 1(h))
The electron ground state of it is a spin triplet. 
Upon optical excitation the NV center shows strong fluorescence allowing it to be observed on an 
individual basis by confocal microscopy. 
The fluorescence intensity of the defect is spin-dependent owing to spin selective relaxation via 
singlet state, which allows optical read out of the single electron spin resonance (ESR)\ \cite{Gru}
and an efficient electron and nuclear initialization at room temperature. 
Microwaves (MW) and radiofrequency (RF) fields are used for coherent manipulation of single electron and 
nuclear spins using conventional ESR/NMR techniques. 
All measurement is carried out at $\sim$20$^{\circ}$C.
Diamond enables for the unique opportunity to control the concentration of paramagnetic nuclear spins. 
The most abundant \C{12} has zero nuclear spin. 
The concentration of \C{13} nuclear spins ($I$=1/2) can be adjusted to the suitable value by controlling 
the isotopic content of \C{13} in the growth medium. 
Two types of synthetic diamonds were used in this study. 
Crystals with 0.35\%, 1.1\%, 8.4\%, 20.7\% \C{13} concentrations were synthesized by a MW plasma-assisted 
homoepitaxial chemical vapor deposition technique (CVD)\ \cite{Isb,Miz} with \C{13}H$_4$ or \C{12}H$_4$ gases. 
The 0.03\% \C{13} and $\sim$100\% \C{13} enriched diamonds were synthesized by high pressure high 
temperature method (HPHT) using getters preventing incorporation of nitrogen into lattice. 
In all samples the concentration of paramagnetic impurity measured by ESR was under detection limit (below 1 ppb) 
except 0.03\% and 100\% \C{13} diamond where the concentration of nitrogen was at 1 ppm level.\ \cite{Nak} 

NV centers with different numbers of \C{13} atoms in the immediate vicinity of the electron spin have been 
investigated in 8.4\% \C{13} diamond. 
Fig. 1(a-d) shows single ESR spectra indicating the hyperfine coupling (HFC) of the 
electron spin to zero, one, two, and three nuclear spins in the first shell (fig.\ 1(h)). 
A magnetic field of $\sim$83 Gauss was oriented along the NV axis ([111]-axis).
To characterize the spin quantum states associated with the transitions in Fig.\ 1(b-d), the spectra were 
simulated by exact diagonalization of the spin Hamiltonian
\begin{equation}
H = g_e\beta_e\tilde{S}B + \tilde{S}DS + (\tilde{S}AI_i-g_n\beta_nI_iB)\label{Hamilonian 1}
\end{equation}
Here the electron spin $S = 1$ and \C{13} nuclear spins in 1st shell are taken into account. 
$\beta_e$ is the Bohr and $\beta_n$ the nuclear magneton, respectively. 
Reported values for zero field splitting (ZFS) parameter ($|D|=2.87$ GHz),\ \cite{Man} isotropic electron and nuclear 
Zeeman g-values ($g_e$=2.0028, $g_n$=1.40483), and HFC parameters of 
$A_{\|}$ = 205 MHz and $A_{\bot}$= 123 MHz\ \cite{Wyk} with angle of 106$^\circ$ between principal axes of ZFS and HFC, 
yield precise fits of the experimental spectra. 
The small splittings in the central signals of the spectra in Fig.\ 1(c,d) are explained by a 2nd 
order perturbation approach.\ \cite{Neu}
The smaller amplitudes in higher frequency are due to absorption of MW by wire on the sample.

In the ESR spectra of 8.4\% \C{13} diamond, basically two types of couplings are immediately visible 
(Fig.\ 1(b-g)), those around 130 MHz originating from first shell \C{13} and those around 14 MHz. 
In $C_{3v}$ symmetry, the number of equivalent atoms in close shells around NV is 3 or 6.
In recent theoretical study,\ \cite{Gal} the 14 MHz splitting are assigned to \C{13} at 
3 and 6 equivalent sites in the 3rd shell (see Video in\ \cite{Vid}.)
From measuring more than 250 individual centers and comparing the probability to find the 14 MHz splitting 
with the one predicted from theory we assign this splitting to nuclei in the 3. shell.\ \cite{Vid}

Individual nuclei in the spin cluster around the electron are addressed via their particular NMR frequency. 
Given the increase in spectral density apparent from Fig.\ 1 one might wonder in how far individual nuclei remain 
addressable.
However, coherent control even in dense spin clusters remains feasible as demonstrated in Fig.\ 2. 
Even in cases where there are three \C{13} in the 1st shell i.e. in total four qubits, Rabi nutations of single 
nuclear spins can be driven by an additional RF as shown in Fig.\ 2(c). 
That is because even multiple RF transition frequencies originating from nuclei at equivalent positions split due to 
higher order HFC contributions as shown above. 
The selectivity is not limited to the relatively large splitting in the 1st shell but can be applied to \C{13} nuclear 
spins in the 3rd shell. 
A spectrum of single NV which has one \C{13} in the 1st shell and one \C{13} in 3rd shell is shown in Fig.\ 1(g). 
We labeled the four nuclear spin states as $\ket{00}, \ket{01}, \ket{10}$, and $\ket{11}$ as shown in Fig.\ 2(a). 
Rabi oscillations between $\ket{00}$ and $\ket{01}$ could be observed as shown in Fig.\ 2(d). 
To obtain a similar Rabi frequency for \C{13} nuclear spins in the 3rd shell as for those in the 1st shell, 
about $1\times10^2$ times higher RF power was necessary. 
This can be interpreted mainly by \textit{hyperfine enhancement},\ \cite{Sch} which predicts that Rabi frequency is almost 
proportional to HFC and the square root of the RF power.
A 10 times smaller HFC in the 3rd shell supports this interpretation.
A figure of merit which characterizes the quality of coherent control was derived by swapping quantum states among 
individual nuclei. 
It was estimated by transferring polarization back to a detectable electron spin state that 82$\pm$5 \% 
of polarization was transferred from $\ket{00}$ to $\ket{01}$ for \C{13} nuclear spins in the 3rd shell.

We are now in a position to engineer simple quantum states in the spin cluster around the electrons spin. 
Bell states $\Phi^{\pm}=1/\sqrt{2}\,(\ket{00}\pm\ket{11})$ and $\Psi^{\pm}=1/\sqrt{2}\,(\ket{01}\pm\ket{10})$ can be 
generated from the four effective nuclear spin states in Fig.\ 2(a).
In the present case they were prepared from two \C{13} spins at the 1st shell. 
The procedure follows previous studies\ \cite{Neu,Meh} and is schematically shown for $\Phi^-$ in Fig.\ 2(k). 
After its generation, $\Phi^-$ shows a free induction decay (FID) which is made visible with an entanglement 
detector sequence (Fig.\ 2(k)). 
In 8.4 \% \C{13} diamond, the free induction decay times $T^*_2$ of $\Psi^-$$(T^*_{2(\Psi)})$ and 
$\Phi^-$$(T^*_{2(\Phi)})$ were estimated to be $22.0\pm3.0 \mu$s and $13.3\pm1.1 \mu$s, (see Fig.\ 2(f,g)) respectively, 
by fitting with $\exp[-(t/T_2^*)^2]\,\cos{(\Delta\omega\,t)}$,\ \cite{Chi} where $\Delta\omega$ is the 
detuned frequency of FID. 
As expected from the view point of decoherence free subspaces,\ \cite{Ros} 
a longer $T^*_{2(\Psi)}$ compared to that of $T^*_{2(\Phi)}$ is observed.

\begin{figure}
\includegraphics{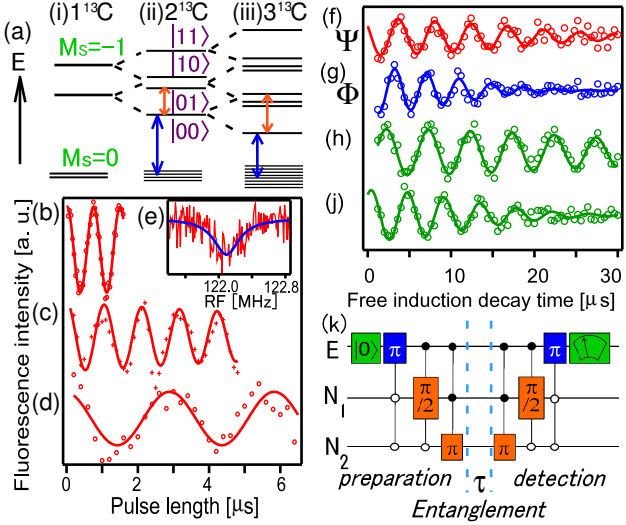}% Here is how to import EPS art[scale=0.48]
\caption{\label{ }(Color online) 
 (a) Energy levels for nuclear spins in M$_S$=-1. 
 Rabi oscillation of single \C{13} in the 1st shell of neighbors around the NV with (b) two \C{13} in the 1st shell, 
 (c) three \C{13} in the 1st shell. 
 (d) Rabi oscillation of single \C{13} in the 3rd shell with one \C{13} in the 1st shell and one \C{13} in the 3rd shell. 
 The pulse sequence is $\pi$(MW)-Rabi(RF)-$\pi$(MW).\ \cite{Fed}
 The ESR transitions of the MW $\pi$ pulse are  those at lowest frequency in  Fig.\ 1 (c,d,g) and are indicated by blue 
 arrows in (a). 
 The NMR transitions of the RF pulse are indicated by orange arrows in (a). The recording the data in (b-d) required 
 about 20 minutes of averaging.
 (e) ENDOR spectrum of \C{13} at 1st shell in (b) with the pulse sequence of $\pi$(MW)-$\pi$(RF)-$\pi$(MW).
 FID of (f) $\Psi^-$, (g) $\Phi^-$, (h) nuclear coherence between $\ket{00}$$\leftrightarrow$$\ket{10}$ and 
 (j) $\ket{00}$$\leftrightarrow$$\ket{01}$. 
 (k) Pulse sequence for $\Phi^-$ generation and detection between two nuclear spins. E and N$_1, _2$ marks the electron 
 and the two nuclear spins, respectively. Spin selective pulses are represented by squares, operating on a target qubit. 
 Vertical lines represent logical connections. 
 The control qubit state $\ket{1}$ and the state $\ket{0}$ are displayed as filled ($\bullet$) and open ($\circ$) circles. 
 For example, ($\circ$) indicates that the pulse is applied to the target qubit if the quantum state of the controlling 
 qubit is $\ket{0}$.}
\end{figure}

The difference among $T_{2(\Psi)}^*$and $T^*_{2(\Phi)}$ is best analyzed when compared 
with $T^*_2$ of a nuclear quantum coherence among states $\ket{00}$$\leftrightarrow$$\ket{10}$ and 
$\ket{00}$$\leftrightarrow$$\ket{01}$. 
Those coherences are labeled as single quantum coherences SQ1 and SQ2, respectively. 
Their $T^*_2$ are measured to be $T^*_{2(SQ1)}$=$41.1\pm3.1 \mu$s and 
$T^*_{2(SQ2)}$=$15.8\pm1.4 \mu$s, respectively (Fig.\ 2(h,j)). 
The difference of $T^*_{2(SQ1)}$ and  $T^*_{2(SQ2)}$ might be caused by a spatially 
inhomogeneous magnetic noise around the defect caused by an inhomogeneous distribution of \C{13} around the two \C{13} 
in the 1st shell. 
Each spin-spin interaction between nuclear spin $k$ surrounding the two \C{13} in the 1st shell with 
quantum numbers $m_{I1}$ and $m_{I2}$ cause oscillation given by 
$\sum_k\exp[-\I(\Delta\omega_1\,m_{I1}+\Delta\omega_2\,m_{I2})t]$.\ \cite{Ref1} 
Here $\Delta\omega_1$ and $\Delta\omega_2$ are spin-spin interaction frequencies of the two \C{13} in the 1st shell 
due to surrounding \C{13} nuclear spins. 
This implies that $T^*_{2(\Psi)}$ and $T^*_{2(\Phi)}$ can be approximated by 
$1/T^*_{2(\Psi)}$=$|1/T^*_{2(SQ1)}-1/T^*_{2(SQ2)}|$ and
$1/T^*_{2(\Phi)}$=$1/T^*_{2(SQ1)}+1/T^*_{2(SQ2)}$, respectively.
Inserting the measured values for $1/T^*_{2(SQ1,2)}$,
the results are $T^*_{2\ \text{calc.}(\Psi)}$=$25.8^{+5.7}_{-4.3} \mu$s and 
$T^*_{2\ \text{calc.}(\Phi)}$=$11.4^{+1.0}_{-0.9} \mu$s, respectively, in good 
correspondence with measured values.

Besides the effect of the nuclear spin bath on individual \C{13} spins the static interaction between the single NV 
electron spin and its environment for different \C{13} concentrations 
was investigated by measuring $T_2^*$ i.e. the inhomogeneous ESR linewidth. 
It was observed that $T_2^*$ increases i.e. the linewidth narrows, with decreasing \C{13} concentration as shown 
in Fig.\ 3. 
In 0.03 \% \C{13} diamond, an extremely long $T_2^*$ of 30 $\mu$s was found (see Fig.\ 3). 
In the low \C{13} concentration region ($\leqq$1.1\%), the linewidth $W$ (full width at half maximum) is derived from 
$T_2^*$ by $W=2\sqrt{ln2}/{\pi}T_2^*$. 
The corresponding 18 kHz linewidth is the narrowest ever observed for an electron spin in a solid material. 
In the high \C{13} concentration region ($\textgreater$1.1\%), the linewidth is derived from fitting the ESR line of 
a single NV with a Gaussian lineshape. 
Average values are plotted as squares in Fig.\ 3(e). 

\begin{figure}
\includegraphics{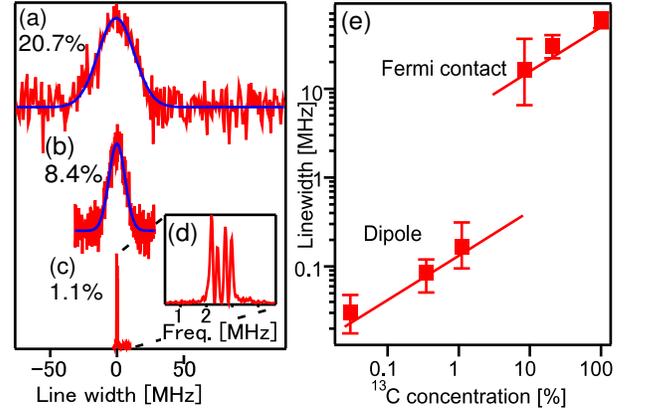}% Here is how to import EPS art[scale=0.56]
 \caption{\label{ }(Color online) 
 ESR spectra of single NV in (a) 20.7 \%, (b) 8.4 \% \C{13} diamond with fitted Gaussian lines (blue). 
 (c) Fourier transformed spectrum of FID of 1.1\% \C{13} diamond shown in (d) on expanded frequency axis. 
 The splitting is due to HFC of distant \C{13} nuclear spins. 
 The hyperfine splitting to N is not visible in this spectrum due to polarization of the N nuclear spin.\ \cite{Neu2}
 (e) Dependence of Inhomogeneous linewidth on \C{13} concentration. 
 The error bars indicates the distributions measured.}
\end{figure}

A likely cause for the inhomogeneous ESR linewidth is HFC to \C{13} nuclear spins. 
In e.g. Si, the dependence of the inhomogeneous linewidth of P donors in $^{29}$Si is well fitted by the 
isotropic HFC ($a_l$) due to Fermi contact interaction with $^{29}$Si nuclear spins with a concentration ($f$), 
%\begin{equation}
$W=2\sqrt{2ln2}[f\sum_l(a_l/2)^2]^{1/2}.$\ \cite{Khn,Abe}
%\end{equation}
The sum runs over all nuclear spin sites $l$. 
In Fig.\ 3(e), the solid line for high \C{13} concentrations is calculated by summing only over all the 9 sites in 
the 3rd shell with $a_l$ = 14 MHz (see above for assignment of sites and HFC constants). 
It should be noted that contributions from \C{13} in the 1st shell were not considered in the linewidth calculations 
because they contribute to an observable splitting but not to the linewidth. 
As seen from Fig.\ 3(e), it fits the experimental results well for high \C{13} concentration.

For lower \C{13} concentration, experimental data deviate from this behavior. 
This is due to the fact that the probability that any \C{13} is located close to the NV center is getting small 
upon reduction of \C{13} concentration. 
Furthermore, the unpaired electron spin density rapidly decreases with distance from the three dangling 
bonds around the vacancy. 
This is known from the HFC parameters\ \cite{Wyk,Gal} which indicates that almost 100 \% spin density is localized on 
the C sites in the 1st and the 3rd shell. 
That is why in this situation the most prominent contribution to the inhomogeneous linewidth is the weaker 
dipole-dipole interaction between electron spin and \C{13} nuclear spin at distant sites. 
The lower line in Fig.\ 3(e) is the linewidth
\begin{equation}
W=\sqrt{(\mu_0\mu_e\mu_ng_eg_n/4\pi{h})^2(3.195\times10^{46}n)},
\end{equation}
 calculated from the 2nd moment\ \cite{Van} with more than 3,000 lattice sites for each \C{13} 
concentration ($n$).
Contributions from \C{13} in the 1st and 2nd shell are not considered.
As seen from Fig.\ 3(e), $W$ fits the experimental results in the low \C{13} concentration ($\leqq$1\%) quite well.
Obviously at low \C{13} concentration the linewidth is dominated by dipole-dipole interaction.

The dephasing time $T_2$ of the electron spin is measured by two pulse Hahn echo decay curves (Fig.\ 4(a-c)). 
We analyzed $T_2$ of the diamond made by CVD and excluded the 0.03 \% and 100 \% \C{13} diamond made 
by HPHT because paramagnetic impurities could not be suppressed in HPHT. 
In the 1.1 \% \C{13} diamond, a $T_2$ of 0.65 ms was found, which is the longest $T_2$ in diamond with natural abundance of 
\C{13} measured so far and for the lower \C{13} concentration of 0.3\% an even longer $T_2$ of 1.8 ms was 
measured.\ \cite{Fed2} 
$T_2$ is found to be inversely proportional to the \C{13} concentration as plotted in Fig.\ 4(d). 

\begin{figure}
 \includegraphics{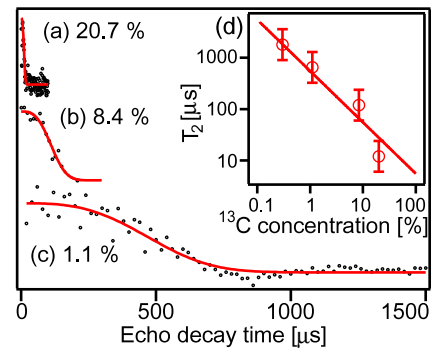}%[scale=0.6]
 \caption{\label{ }(Color online) Echo decays of electron spin in (a) 20.7 \%, (b) 8.4 \%, and (c) 1.1\% \C{13} 
 diamond. 
 The MW pulse sequence is $\pi/2$-$\tau$-$\pi$-$\tau$-$\pi/2$ where $\tau$ is delay\ \cite{Chi}.
 Red lines are curves fitted with $exp[-(t/T_2)^3$]. (d) Plot of $T_2$ over \C{13} concentration $n$. 
 The solid line is fitted with a 1/$n$ dependence.}
\end{figure}

In a theoretical analysis of $T_2$ by the disjoint cluster approach,\ \cite{Maz} the relationship of 
$T_2\sim(\bar{C}A_c)^{-1/2}$ is proposed, where $\bar{C}$ is the averaged nuclear-nuclear dipolar interaction in the bath and 
$A_c$ is some characteristic value for the electron-nuclear dipolar interaction. 
Since both interactions scale linearly in \C{13} concentration ($n$), $T_2$ decreases approximately as 1/$n$ in this 
model. 
The fitted line to the data shown in fig. 4 (d) supports this inverse proportionality and fits our data. 
Note that our data also fit the values calculated in \cite{Maz} within 30\% deviation.

In conclusion, coherent control of up to three individual nuclei in a dense nuclear spin cluster is demonstrated. 
The \C{13} concentration dependence of $T_2^*$ and $T_2$ of electron spin point towards \C{13} nuclei as the main 
cause for dephasing in otherwise clean diamond.
The correspondence with the theoretical line of $T_2$ \cite{Maz} is very important to elucidate the dephasing 
mechanism and to make $T_2$ longer for quantum information devices \cite{Jia} and magnetometry. \cite{Gop,Ma2}
Furthermore, the results show that the threshold ($\sim$10$^4$ operation) for quantum error-correction 
schemes \cite{Aws} can be exceeded even in \C{13} enriched diamond at room temperature with typical 
single-qubit flip of several ns.

This work is supported by the EU (QAP, EQUIND, NANO4DRUGS, NEDQIT), DFG (SFB/TR21, FOR730), JST-DFG 
program, KAKENHI (20760006) and the Landesstiftung BW. VJ acknowledges 
support by the Humboldt Stiftung. We thank Dr. H. Kanda for providing 100\% \C{13} diamond.

%\newpage %Just because of unusual number of tables stacked at end
\bibliography{apssamp}% Produces the bibliography via BibTeX.

\end{document}